\documentclass[onecolumn,amsmath,amssymb]{qm2008_abs}
\usepackage{graphicx}
\usepackage{dcolumn}
\usepackage{bm}
\begin{document}

\title{{\Large Properties of the $\phi$ meson at high temperatures and densities }}

\bigskip
\bigskip
\author{\large G. Vujanovic}\email{gojkov@physics.mcgill.ca}\affiliation{Physics Department, McGill University, Montreal, QC, H3A 2T8 Canada}
\author{\large J. Ruppert}\affiliation{Physics Department, McGill University, Montreal, QC, H3A 2T8 Canada}\affiliation{Institut f\"ur Theoretische Physik, J.W. Goethe University Frankfurt, Max-von-Laue-Str. 1, D-60438 Frankfurt am Main, Germany}
\author{\large C. Gale}\affiliation{Physics Department, McGill University, Montreal, QC, H3A 2T8 Canada}
\bigskip
\bigskip

\begin{abstract}
\leftskip1.0cm
\rightskip1.0cm
We calculate the spectral density of the $\phi$ meson in a hot bath of nucleons and pions using a general formalism relating self-energy to the forward scattering amplitude (FSA) \cite{eletsky-belkacem-kapusta,martell-ellis}. In order to describe the low energy FSA, we use experimental data along with a background term. For the high energy FSA, a Regge parameterization is employed. We verify the resulting FSA using dispersion techniques. We find that the position of the peak of the spectral density is slightly shifted from its vacuum position and that its width is considerably increased. The width of the spectral density at a temperature of 150 MeV and at normal nuclear density is more than 90 MeV.
\end{abstract}

\maketitle

\section{Introduction}

Electromagnetic decays of vector mesons are a good probe of the hot and dense state created after a nucleus-nucleus collision. Indeed, the emitted lepton pairs, which carry the spectral information of the vector meson at the moment of the decay, undergo a negligible distortion as they travel through the dense medium \cite{kapusta-gale}. This fact can in turn tell us about properties of the medium.  

\section{Theory}
We consider interactions with pions ($\pi$) and nucleons ($N$) as scatterings of the $\phi$ meson that contribute to its self-energy. In this approach the forward scattering amplitude (FSA) is composed of a low energy and a high energy part. The latter is described by Regge parameterization whereas former encompasses two terms, namely a resonance term modeled by Breit-Wigner functions and background term. In the center of mass (c.m.) frame the low energy FSA is written as:
\begin{equation}
f^{\rm c.m.}_{\phi a}(s) = \frac{1}{2q_{\rm c.m.}}\sum_{R} W^{R}_{\phi a}\frac{\Gamma_{R\rightarrow \phi a}}{M_{R} - \sqrt{s} - \frac{1}{2}i\Gamma_{R}} - \frac{q_{\rm c.m.}}{4 \pi s} \frac{1+\exp(-i \pi \alpha_P)}{\sin(\pi \alpha_P)} r_{\phi a}^{P} s^{\alpha_P} \label{eq:f_low}
\end{equation}
Here the sum ranges over all Breit-Wigner resonances that decay into the $\phi$ meson and the particle $a$ which is either a nucleon or a pion. We now briefly explain the different variables involved. The mass of the resonance $R$ is $\mathrm{M_R}$ and its total width is $\mathrm{\Gamma_R}$. $s$ is the usual Mandelstam variable and the magnitude of the c.m. momentum is given by:
\begin{equation}
q_{\rm c.m.} = \frac{1}{2} \frac{ \sqrt{\left[ s - \left( m_\phi + m_a \right)^2 \right] \left[ s - \left( m_\phi - m_a \right)^2 \right]} }{ \sqrt{s} }\label{eq:qcm}
\end{equation}
where $m_a=m_N,m_\pi$. $W^{R}_{\phi a}$ are statistical spin/isospin averaging factors:
\begin{equation}
W^{R}_{\phi a} = \frac{(2s_R + 1)}{(2s_{\phi}+1)(2s_a+1)} \frac{(2t_R + 1)}{(2t_{\phi} + 1)(2t_a + 1)}
\end{equation}
where $s_i$ (with $i$ generic) is the spin of particle $i$ and $t_i$ is the isospin of that particle.

The effective width $\Gamma_{R\rightarrow \phi a}$ takes the form \cite{eletsky-belkacem-kapusta}:
\begin{eqnarray}
\Gamma_{R \rightarrow \phi a} = \left\{ \begin{array}{ll}
\Gamma_R B_{R \rightarrow \phi a} \left( \frac{q_{\rm c.m.}}{q^R_{c.m.}} \right)^{2l+1} & \textrm{$q_{\rm c.m.}$ $\leq$ $q^{R}_{\rm c.m.}$}\\
\Gamma_R B_{R \rightarrow \phi a} & \textrm{$q_{\rm c.m.}$ $\geq$ $q^{R}_{\rm c.m.}$} \\
\end{array} \right.\label{eq:effective-width-threshold-n}
\end{eqnarray}
where $\Gamma_{R}$ is the total width of the resonance $R$, $B_{R \rightarrow \phi a}$ is the branching ratio of the decay $R \rightarrow \phi a$, $q^{R}_{\rm c.m.} =  \frac{1}{2} \frac{ \sqrt{\left[ M^{2}_{R} - \left( m_\phi + m_a \right)^2 \right] \left[ M^{2}_{R} - \left( m_\phi - m_a \right)^2 \right]} }{ M_{R} }$, and $l$ is the minimal relative angular momentum between $\phi$ and $a$. The case of the pions is a bit more complicated for the $l=0$ case however, as Adler's theorem should be fulfilled. That is, the pion scattering amplitude on any hadronic target vanishes when $q_{\rm c.m.} \rightarrow 0$ in the limit of massless pions \cite{eletsky-belkacem-kapusta}. Consequently, the effective width for $l=0$ now must be rewritten as:  
\begin{eqnarray}
\Gamma_{R \rightarrow \phi \pi} = \left\{ \begin{array}{ll}
\Gamma_R B_{R \rightarrow \phi \pi} \left( \frac{s-m_{\phi}^2-m_{\pi}^2}{s_0-m_{\phi}^2-m_{\pi}^2} \right)^2 & \textrm{$s$ $\leq$ $s_0$}\\
\Gamma_R B_{R \rightarrow \phi \pi} & \textrm{$s$ $\geq$ $s_0$} \\
\end{array} \right. \label{eq:adler-theorem-phi}
\end{eqnarray}
with $s_0=\left( m_{\phi} + m_{\pi} + m_{\rho} \right)^2$. Finally, the high energy FSA is well described by a Regge parametrization of the form:
\begin{equation}
f^{\rm c.m.}_{\phi a} \left(s\right)= -\frac{q_{\rm c.m.}}{4 \pi s} \sum_{i} \left[ \frac{1+\exp(-i \pi \alpha_i)}{\sin(\pi \alpha_i)} \right] r_{\phi a}^{i} s^{\alpha_i}. \label{eq:f_high}
\end{equation}
where $a$ can be a nucleon or a pion as mentioned before. The variables $r_{\phi a}^{i} $ and $\alpha_i$ are Regge residues and intercepts which are determined experimentally \cite{pdg}. 

\section{Forward scattering amplitude}

\begin{figure}[!ht]
\begin{center}
\begin{tabular}{c c}
\includegraphics[scale=0.31]{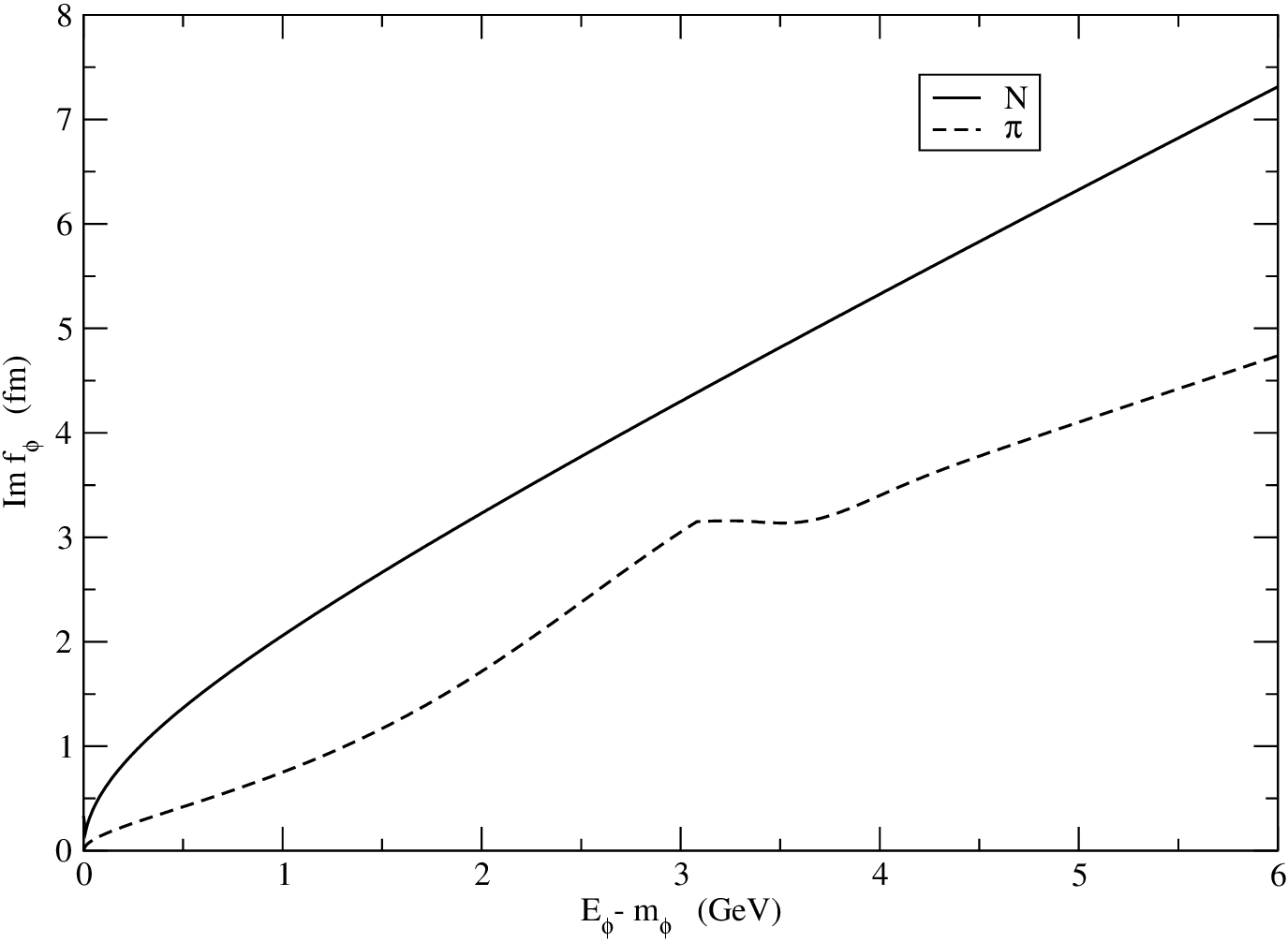} & \hspace{2mm}  \includegraphics[scale=0.31]{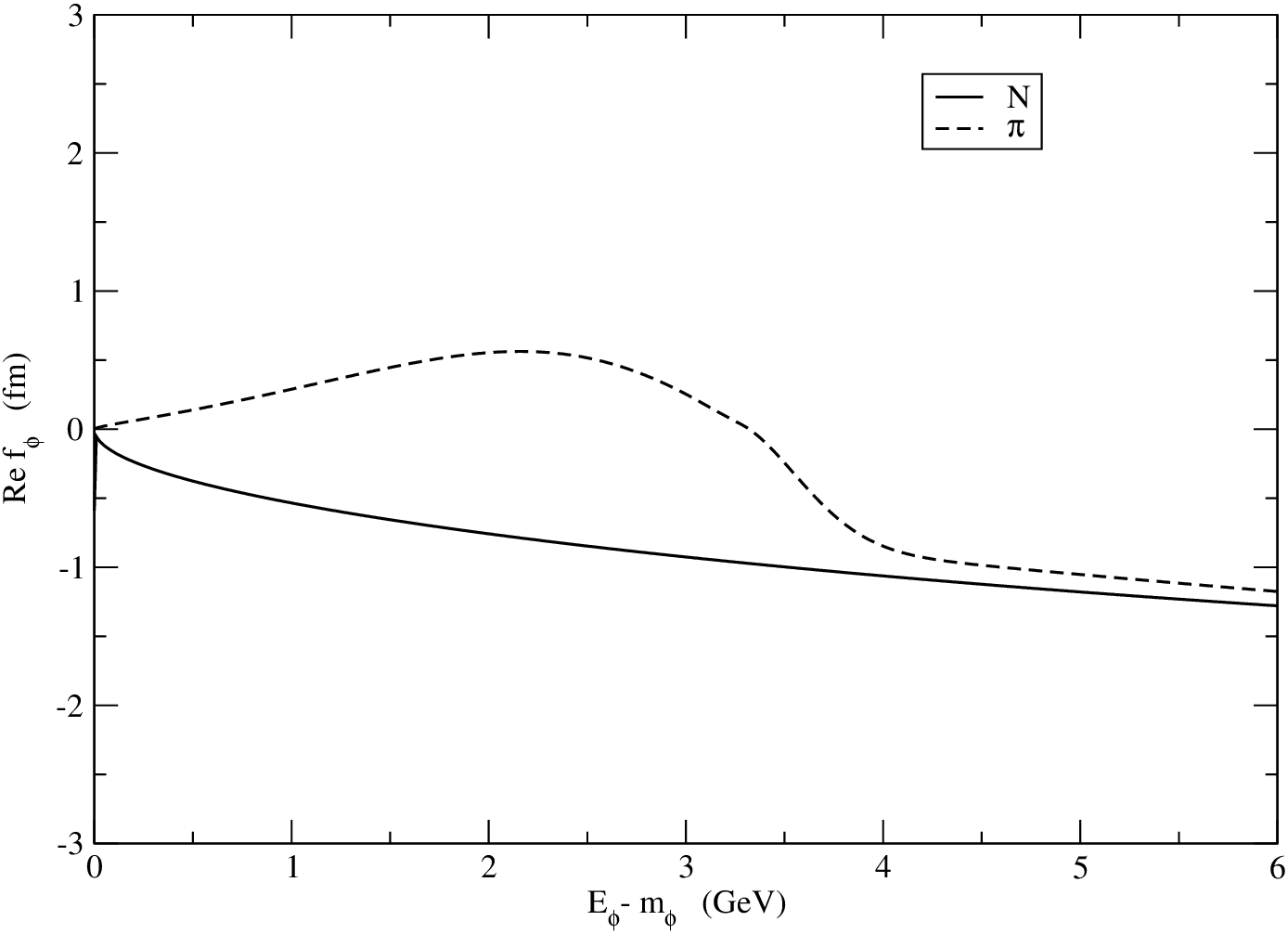}\\
(a) & (b)
\end{tabular}
\end{center}
\caption{(a) The imaginary and (b) the real part of the FSA for both $\phi N$ (solid line) scattering and $\phi \pi$ scattering (dashed line).}\label{pic:re_im_f}
\end{figure}

The real and imaginary parts of the FSA are plotted in Fig.\ref{pic:re_im_f}. These plots are made relative to the rest frame of the heath bath. 
We use a dispersion relation in order to ensure that the matching between the low energy and the high energy of the FSA is correctly done. 
\begin{equation}
{\rm Re}\left[f^{\rm total}_{\phi a} \left(E_\phi \right) \right] = {\rm Re} \left[f^{\rm total}_{\phi a} \left( 0 \right) \right] + \frac{2E^{2}_{\phi}}{\pi} \mathrm{P.V.} \int^{\infty}_{m_\phi} \frac{{\rm Im} \left[ f^{\rm total}_{\phi a}\left( E^{\prime}\right) \right]  d E^{\prime}}{E^{\prime}\left( E^{\prime}+E_{\phi}\right) \left( E^{\prime} - E_{\phi}\right)}. 
\end{equation}
If the difference between the real part of the FSA and the one calculated from the dispersion relation is a constant, then the matching is perfect. Of course, it should not be expected that our phenomenological approximations exactly obey the constraints that follow from the analytic properties of the FSA \cite{eletsky-belkacem-kapusta} and therefore, some deviations should occur.         
\vspace{0.5cm}
\begin{figure}[!ht]
\begin{center}
\includegraphics[scale=0.31]{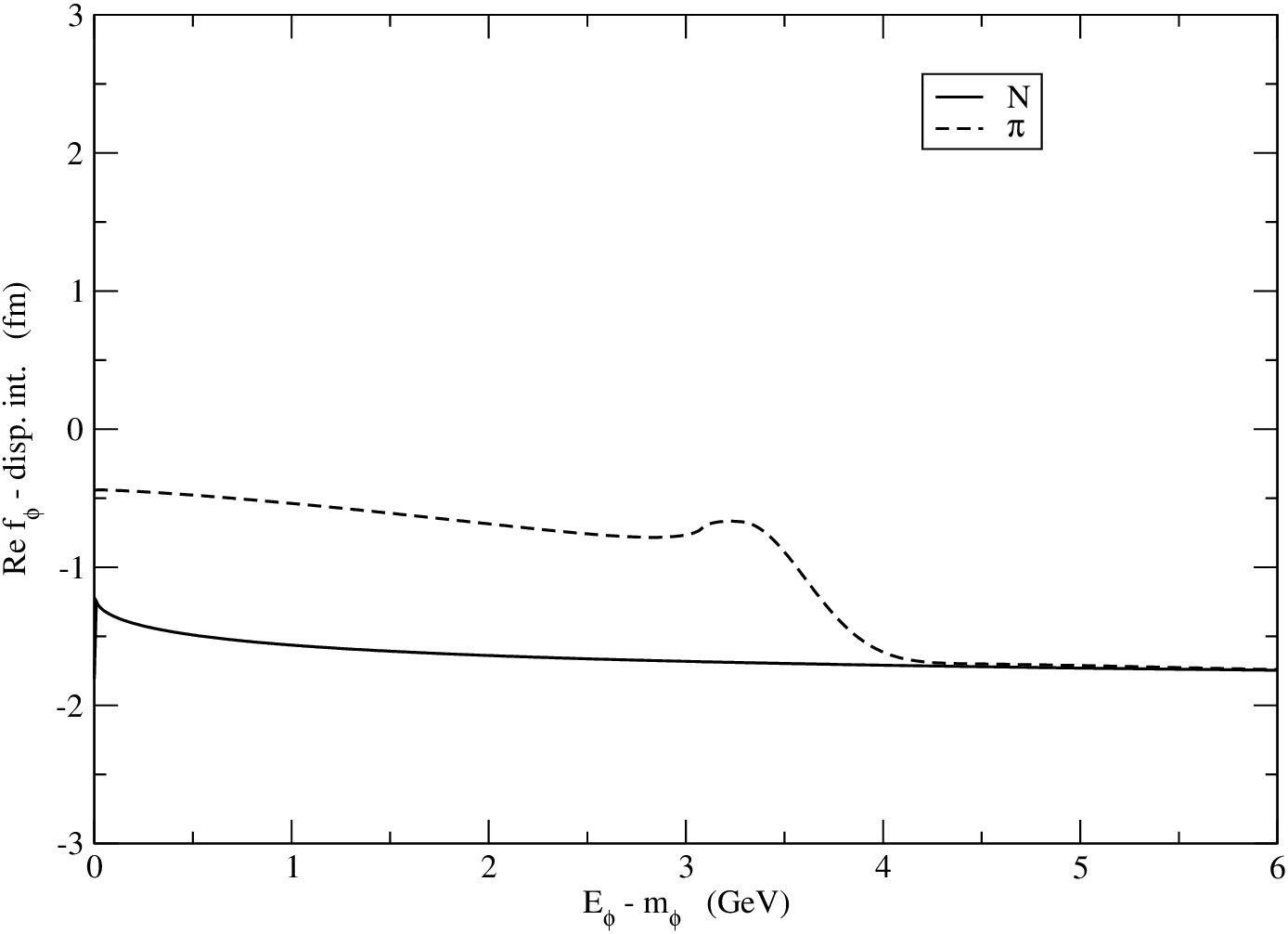}
\end{center}
\caption{Difference between the real part of the FSA in Fig. \ref{pic:re_im_f} (b) and the one calculated one from the dispersion relation.}\label{pic:dispersion}
\end{figure}

\section{Self-energy of the $\phi$ meson}

\begin{figure}[!ht]
\begin{center}
\begin{tabular}{c c}
\includegraphics[scale=0.31]{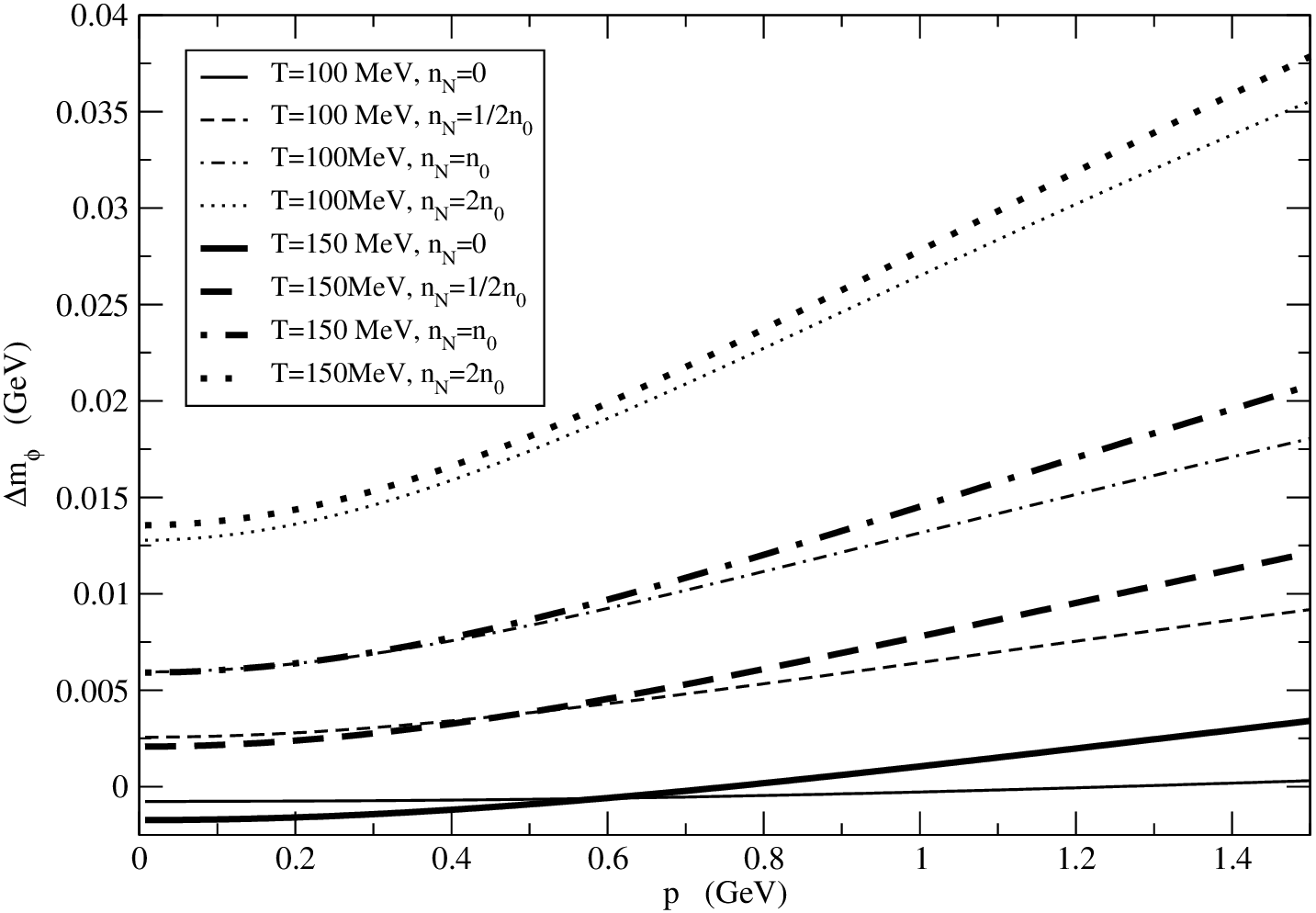} & \hspace{1mm} \includegraphics[scale=0.31]{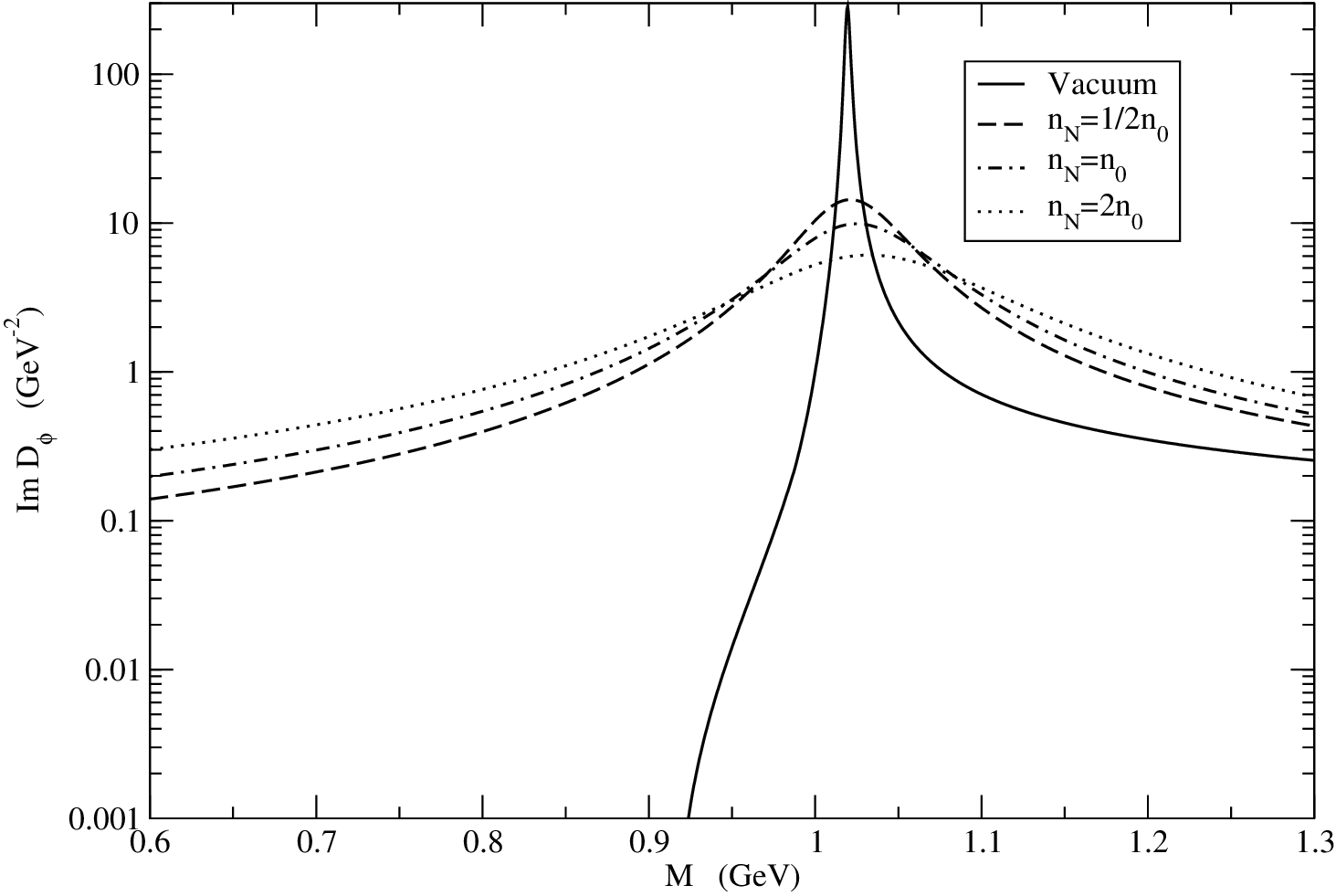}\\
(a) & (b)
\end{tabular}
\end{center}
\caption{(a) Change in mass of the $\phi$ meson as a function of $p$, $n_N$ and $T$. The results presented include four different nucleon densities (namely $0$, $\frac{1}{2}n_0$, $n_0$, $2n_0$ with $n_0=0.16$ $\mathrm{nucleons}/\mathrm{fm}^3$) and two temperatures (i.e. $T=100$ MeV, $T=150$ MeV). (b) The imaginary part of the $\phi$ meson propagator as a function of invariant mass $M$, for a 3-momentum of 0.3 GeV and a temperature of 150 MeV. The results include three nucleon densities (namely $\frac{1}{2}n_0$, $n_0$ and $2n_0$). The vacuum contribution is also displayed. }\label{pic:delta_m-im_prop}
\end{figure}

The forward scattering amplitude is related to the in-medium self-energy of the $\phi$ meson by \cite{eletsky-belkacem-kapusta}:
\begin{equation}
\Pi_{\phi a} \left(E,p\right)= -4\pi \int \frac{d^3 k}{\left( 2\pi \right)^3}n_a\left(\omega\right)\frac{\sqrt{s}}{\omega}f^{\rm total, c.m.}_{\phi a}\left(s\right) 
\end{equation}
where $E$ and $p$ are the energy and momentum of the $\phi$ meson, $\omega^2=m^2_a+k^2$, and $n_a$ is either a Fermi-Dirac or a Bose-Einstein distribution depending on the nature of $a$. The total self-energy is given by:
\begin{equation}
\Pi^{\rm total}_\phi (E, p)= \Pi^{\rm vac}_\phi \left( M \right) + \Pi_{\phi \pi} \left( p \right) + \Pi_{\phi N}\left(p\right)
\end{equation}
The vacuum self-energy includes the charged and neutral $\phi\rightarrow K \bar{K}$ channels as well as $\phi\rightarrow\rho\pi$ channel. The mass shift of the $\phi$ meson due to its interaction with the medium is:
\begin{equation}
\Delta m_\phi \left( p \right) = \sqrt{m^2_\phi+ {\rm Re} \left[\Pi^{\rm total}_\phi \left( p \right)  \right]}- m_\phi
\end{equation}
The mass shift for the $\phi$ meson is presented in Fig. \ref{pic:delta_m-im_prop} (a). The rate of dilepton production is proportional to the imaginary part of the propagator due to vector meson dominance \cite{kapusta-gale}.
\begin{equation}
E_{+} E_{-} \frac{dR}{d^3p_{+}d^3p_{-}} \propto \frac{{\rm Im} \left[ \Pi^{\rm total}_\phi \right]}{\left\{M^2-m^2_\phi - {\rm Re} \left[ \Pi^{total}_\phi \right] \right\}^2+\left\{ {\rm Im} \left[ \Pi^{\rm total}_\phi \right] \right\}^2}
\end{equation}
The imaginary part of the propagator is presented in Fig. \ref{pic:delta_m-im_prop} (b) where we clearly see the width broadening of the $\phi$ meson. 
\vspace{0.5cm}
\section{Summary}
The results show that the mass shift overall is small, reaching 38 MeV at most. There is a significant broadening of the width of the $\phi$ meson however. At normal nucleon density and temperature of 150MeV, the width  more than 90 MeV. Our results are important for understanding possible in-medium modifications of the $\phi$ meson strongly interacting environments. Studies to estimate phenomenological implications are in progress \cite{paper-in-progress}. \\
\\

\end{document}